\documentclass[11pt]{article}

\usepackage{jcappub}
\usepackage{amsmath,amssymb,amsfonts,tikz}
\usepackage{amsthm}
\usepackage{relsize}
\usepackage{float}
\usepackage{placeins}
\usepackage{tensor}

\usetikzlibrary{decorations.pathreplacing}
\usetikzlibrary{decorations.pathmorphing}
\usetikzlibrary{arrows}
\usepackage{esint}
\usepackage{graphicx,tikz}
\usepackage{adjustbox}
\usepackage[squaren,Gray]{SIunits}
\usepackage{epstopdf}
\usepackage{listings}
\usepackage{xcolor}
\colorlet{shadecolor}{gray!20}
\usepackage{mathtools}
\usepackage{physics}
\usepackage{caption}
\usepackage{subcaption}
\usepackage[margin=1in]{geometry}
\usepackage[title,titletoc,toc]{appendix}
\usepackage{mdframed}
\usepackage{bigints}
\usepackage{multirow}

\usepackage{mathrsfs}
\usepackage{setspace}
\usepackage{lineno,hyperref}
\usepackage{soul}
\usepackage{titlesec}
\usepackage{multirow}
\usepackage{placeins}
\numberwithin{equation}{section}
	
\newcommand{\R}{\mathbb{R}}

\newcommand{\Ll}{\mathcal{L}}

\renewcommand{\vec}[1]{\mathbf{#1}}

\renewcommand{\d}{\,\mathrm{d}}
\newcommand{\diag}{\operatorname{diag}}

\newcommand{\del}{\partial}
\renewcommand{\so}{\text{SO}}
\newcommand{\iso}{\text{ISO}}
\newcommand{\dbphi}{\dot{\bar\varphi}}

\newcommand*\pFq[2]{{}_{#1}F_{#2}}

\definecolor{light-gray}{gray}{0.80}
\newcommand{\highlight}[1]{\colorbox{light-gray}{\ensuremath{#1}}}

\newtheorem{theorem}{Theorem}[section]
\newtheorem*{theorem*}{Theorem}

\newtheorem*{Remark*}{Remark}
\newtheorem{claim}{Claim}[theorem]

\newcommand{\be}{\begin{equation}}
\newcommand{\ee}{\end{equation}}
	
\title{Symmetric Scalars}
\author{Tanguy Grall, Sadra Jazayeri and Enrico Pajer}
\affiliation{Center for Theoretical Cosmology, Department of Applied Mathematics and Theoretical Physics, University of Cambridge, Wilberforce Road, Cambridge CB3 0WA, UK}

\abstract{
We provide a complete classification of Poincar\'e-invariant scalar field theories with an enlarged set of classical symmetries to leading order in derivatives, namely for the so-called $P(X,\phi)$ theories, in two or more spacetime dimensions. We find only three possibilities: Dirac-Born-Infeld, Cuscuton and Scaling theories. The latter two classes of actions involve an arbitrary function of the scalar field. As an application, we use the scaling symmetry to derive an infinite set of constraints on the Wilsonian coefficients of the low-energy Effective Field Theory. Furthermore, we study the extension of these results to cosmological (FLRW) and (Anti-)de Sitter spacetimes. We find in particular that the Cuscuton action has a generic set of symmetries around any background spacetime that possesses Killing vector fields, while the DBI actions have well-known analogues that we summarize explicitly. 
}
\begin{document}

\maketitle
\flushbottom




\section{Introduction} 

Symmetries, namely the invariance of physical laws as we change our point of view, play a major role in the way we describe natural phenomena. In Poincar\'e invariant quantum field theories, linearly-realized continuous symmetries, which act trivially on the vacuum state, are highly restricted by the results of Coleman and Mandula \cite{Coleman:1967ad} and by the supersymmetric generalization of Haag, $ \L  $opusza\'nski and Sohnius \cite{Haag:1974qh}. For example, no bosonic spacetime symmetries can act non-trivially on scattering amplitudes. On the other hand, when Poincar\'e or additional symmetries are spontaneously broken by the vacuum state and therefore non-linearly realized many more possibilities arise. Charting this wider landscape of symmetric theories is particularly important for cosmology and condensed matter physics, where Poincar\'e symmetries are necessarily broken.

When Poincar\'e symmetries are unbroken, additional non-linearly realized symmetries can be classified by studying the soft limits of scattering amplitudes \cite{,Adler:1964um,Low:2014nga,Cheung:2014dqa,Cheung:2016drk,Padilla:2016mno,Bogers:2018zeg,Roest:2019oiw,Roest:2019dxy}. But in cosmology time translations are spontaneously broken by the expansion of the universe and so the notion of asymptotic states used to define scattering amplitudes is unclear. So, despite the fact that symmetries are routinely used to model cosmological phenomena, few results are known about the most generic set of symmetries that are allowed for a given set of fields. 

In this work, we make a modest step forward in this direction by providing a complete classification of all Poincar\'e invariant theories of a single scalar field that admit additional continuous symmetries to leading order in derivatives. Our results include cases where either the Poincar\'e or the additional symmetries are spontaneously broken. We derive this classification by brute force: we write down the most generic Lagrangian and try adding more and more symmetries until there are no more free functions or couplings to be specified. We use the commutators of these new symmetries with translations and Lorentz transformations as organizing principle for the calculation. This work thus represents an extension of \cite{Pajer:2018egx}, where only shift-symmetric actions were classified.

Our final results are summarized in Table \ref{tab:all_algebras_mink}. Only three classes of theories exist that enjoy additional (continuous) symmetries: scalar Dirac-Born-Infeld (DBI) \cite{Alishahiha:2004eh}, Cuscuton \cite{Afshordi:2006ad} and Scaling theories (that are invariant under dilations). For DBI, the symmetries are so strong that only a single coupling constant in the potential remains unspecified. For Scaling and Cuscuton theories instead, the most generic Lagrangian contains a free function, which can be further constrained by adding even further symmetries. Some reader might be surprised that a single scalar can realize vectorial and tensorial symmetries. The reason is that all the symmetries we find are spacetime symmetries, i.e.~they do not commute with the Poincar\'e group. Then Goldstone theorem does not apply and one can have less Goldstone bosons than broken generators (see e.g. the nice discussion in \cite{Low:2001bw}). 

The remainder of the paper is structured as follows. In Sec.~\ref{sec:minkowksi_background}, we introduce our methodology and derive the full classification of $P(X,\phi)$ actions along with the corresponding symmetry algebras. With these results in hand we can then study the behaviour of each theory around non-trivial background field profiles that spontaneously break spacetime symmetries. In particular with cosmology in mind, we want to investigate the dynamics of perturbations around homogeneous but time-dependent backgrounds. DBI and Cuscuton theories have already been extensively studied in the context of cosmology (see e.g.~\cite{Alishahiha:2004eh,Hinterbichler:2012fr,Renaux-Petel:2013ppa,Creminelli:2013xfa} and \cite{Afshordi:2007yx,Afshordi:2009tt,Gomes:2017tzd,Ito:2019ztb,Ito:2019fie} respectively). And while scaling symmetry has also attracted much attention in cosmological model building (see \cite{Wetterich:1987fm,Quiros:2014wda,Padilla:2013jza,Ferreira:2016vsc,Kannike:2016wuy,Ferreira:2016wem,Rubio:2017gty,Ferreira:2018qss} for an incomplete list of references), to our knowledge the dynamics of perturbations around non trivial backgrounds of a single minimally coupled scaling-invariant scalar field have not been worked out. We therefore investigate this, together with the consequences of scaling symmetry for the EFT of Inflation \cite{Cheung:2007st} in Sec.~\ref{sec:scaling_theories}. We find that the symmetry generates an infinite set of recursion relations among the Wilsonian coefficients of the EFT and their time dependence. In Sec.~\ref{sec:(A)dS_backgrounds} we extend our scope and discuss what happens to these symmetric theories when minimally coupled to curved background spacetimes focussing on de Sitter (dS), Anti-de Sitter (AdS) and FLRW spacetimes. Remarkably all three theories exist with analogous symmetries on these spacetimes. Some of these curved spacetime extensions were already known in the literature while others are new. Finally, we conclude and discuss avenues for future work in Sec.~\ref{sec:discussion}.
\section*{Conventions} 
\label{sec:conventions}
 We work with mostly plus signature for the D-dimensional Minkowski metric $\eta_{\mu \nu}=\diag(-1,1,\dots,1)$. Greek indices $\mu, \nu \dots$ run from 0 to $d=D-1$, capital Latin indices $A,B,\dots$ from $-1$ to $D$ and lower case Latin indices $a,b,\dots$ from $-1$ to $d$. Our basis for the Conformal algebra is defined from the $\so(2,D)$ generators $M_{AB}$ as
$P_{\mu}=M_{-1 \mu}+M_{D \mu}$, $K_{\mu}=M_{-1 \mu}-M_{D \mu}$ and $\hat D=M_{D-1}$, where the hat on $ \hat D$ is used to distinguish the generator from the number of spacetime dimensions. The commutation relations read:
\begin{align}
	\label{eq:conventions_conf_algebra}
	\comm{P_{\mu}}{\hat D}&=-P_{\mu}\,, && \comm{K_{\mu}}{\hat D}= K_{\mu}\,,\\ \nonumber
	\comm{K_{\mu}}{P_{\nu}}&=2\left(M_{\mu \nu}-\eta_{\mu \nu} \hat D\right) \,,&& \comm{K_{\mu}}{K_{\nu}}=0\,,
\end{align}
and $P_{\mu}, M_{\mu \nu}$ satisfy the commutation relations of the Poincar\'e subalgebra $\iso(1,d)$ of $\so(2,D)$. We also denote the radius of $D $-dimensional dS $L_D$ and that of $D $-dimensional AdS $R_D$.

\label{sec:introduction}
\begin{table}[h!]
\centering
\resizebox{\columnwidth}{!}{
\begin{tabular}{| c | c | c | c | }
\hline
\multicolumn{4}{|c|}{Minkowski Background} \\
\hline
 Theory & Lagrangian & Generators & Algebra \\
\hline
&&&\\[0.002ex]
\multirow{2}{*}{DBI} & $ \sqrt{1\pm X} + \lambda\, \phi $& $Q, \hat{D}, A^{(S)}_{\mu}$ & $ \text{ISO}(D,1)$ (non-linearly realized)\\
&&&\\[0.002ex]
& $ e^{-D\phi/R_D} \left( \sqrt{1\pm e^{2\phi/R_D}X}+ \lambda  \right)$& $\hat{D}, A^{(NS)}_{\mu}$ & $\text{SO}(D,2)$ (non-linearly realized)\\
&&&\\[0.002ex]
\hline
&&&\\[0.002ex]
\multirow{4}{*}{Scaling} & 
$\phi^{\frac{D}{\Delta}}h\left(\phi^{-2 \frac{\Delta+1}{\Delta}}X\right)$& $\hat{D}$ & $\R^D\rtimes(\text{SO}(d,1)\times \R)$ \\
&&&\\[0.002ex]
&$X^{\alpha}$& $Q, \hat{D}$ & $U(1)\rtimes (\R^D\rtimes(\text{SO}(d,1)\times \R))$ \\
&&&\\[0.002ex]
& $X^{D/2}$& $Q, \hat{D}, K^{(S)}_{\mu}$ & $U(1)\times \text{SO}(D,2)$ \\
&&&\\[0.002ex]
& $
\begin{cases}
		\mathlarger{\int} \d^D x \left(X - \lambda\phi^{\frac{2D}{D-2}}\right) & D\neq2 \\
		\mathlarger{\int} \d^D x \left(\frac{X}{\phi^2}- \lambda\phi^{2/\Delta}\right) & D=2\,
	\end{cases}$& $\hat{D}, K^{(NS)}_{\mu}$ & $\text{SO}(D,2)$ \\
&&&\\[0.002ex]
\hline
&&&\\[0.002ex]
\multirow{3}{*}{Cuscuton}& $\sqrt{X}-V(\phi)$& $V_{\mu}, T_{\mu \nu}$ & $\R^{\infty}_V\rtimes(\R^{\infty}_T\rtimes \text{ISO}(d,1))$   \\
&&&\\[0.002ex]
& $\sqrt{X}-\lambda\, \phi^{\frac{D}{d}}$& $\hat{D}, V_{\mu}, T_{\mu \nu}$ & $\R^{\infty}_V\rtimes(\R^{\infty}_T\rtimes (\R^D\rtimes(\text{SO}(d,1)\times \R)))$\\
&&&\\[0.002ex]
& $\sqrt{X}$& $Q, W, V_{\mu}, T_{\mu \nu}$ & $\R^{\infty}_V\rtimes (\R^{\infty}_T\rtimes (U(1)\rtimes (\R^D\rtimes(\text{SO}(d,1)\times \R_W^{\infty}))))$ \\
&&&\\[0.002ex]
\hline
\end{tabular}
}
\caption{All possible extensions of the Poincar\'e  algebra in $D=d+1$ dimensions for $P(X,\phi)$ theories and $D\geq 2$ on Minkowksi spacetime. This table includes both the results of this paper and those of \cite{Pajer:2018egx}. The dimensionless coupling constants $\lambda$ are arbitrary and in particular can be set to zero and $\alpha=D/2(1+\Delta)$.}
\label{tab:all_algebras_mink}
\end{table}

%
%
\section{Classification of symmetric theories on Minkowski spacetime} 
\label{sec:minkowksi_background}

After reviewing our methodology in Sec.~\ref{ss:meth}, we derive all possible symmetric Lagrangians organizing the exploration in terms of the commutators of the new symmetry generators with spacetime translations.


\subsection{Methodology} \label{ss:meth}

We aim to classify all field theories for a single scalar $  \phi $ of the form
\begin{equation}
	\label{eq:P(X,phi) action}
	S=\int\d^Dx \,P(X,\phi)\, ,
\end{equation}
with $X=-\eta^{\mu \nu}\partial_\mu\phi\partial_\nu\phi$ that admit extra symmetries, namely symmetries that do not commute with the Poincar\'e symmetries. To this end, we follow the strategy outlined in \cite{Pajer:2018egx}, which we summarize in the following. To begin with, we consider consistent Lie algebras that extend the Poincar\'e algebra. This requires that new symmetry generators $  S $ should fall into representations\footnote{We will work with reducible representations of the Lorentz group.} of the Lorentz group---they have to be covariant tensors. We also assume, in the same vein as e.g.~\cite{Coleman:1967ad,Klein:2018ylk}, that the symmetry transformation of the scalar field under the action of $S$ is a polynomial in $x^\mu$: 
\begin{equation}
\delta_{S^N_m}\phi=\sum_{n=0}^N\,x^{\mu_1}\,\dots\,x^{\mu_n}\,f_{\mu_1 \dots\mu_n\mu_{n+1}\,\dots\,\mu_{n+m}}\,,
\end{equation}
where the upper index $  N $ on $  S^{N}_{m} $ indicates the order of the polynomial while the lower index $  m $ refers to the number of Lorentz indices, $  S_{m}=S_{\mu_{1}\dots \mu_{m}} $. In the following, we will assume that $  N $ is finite, as in \cite{Pajer:2018egx}. The coefficients $f_{\mu_1\dots}$ generically may depend on $\phi$ and its derivatives, but we shortly demonstrate that having conserved Noether currents enforces $f_{\mu_1\dots}$ to depend only on $\phi$ and $\partial_\mu\,\phi$. Notice that commuting $S^N_m$ with the translations $P^\mu$ - with $\delta_{P^\mu}\phi=-\partial_\mu\phi$ - gives another symmetry of degree $N-1$\footnote{See Appendix \ref{app:definition_comm} for the  definition of the commutator we make use of throughout the paper.}:
\be
\label{eq:ladder_equation}
[S^{N}_m,P^\mu]=S^{N-1}_{m+1}\,.
\ee
In general, pairs of the Lorentz indices of $  S_{m+1}^{N} $ may be those of the Minkowski metric $  \eta_{\mu\nu} $, while others may be associated with non-c-number operators. At the level of one derivative per field, which we will always assume in this paper, there is no non-vanishing contraction involving the Levi-Civita symbol $  \epsilon_{\mu\nu\rho\sigma\dots} $. By closure of the algebra, commutators of symmetry generators must be also symmetry generators. This property will help us classify all possible degree $N$ symmetries based on the knowledge of degree $N-1$ symmetries in an inductive manner. 

After having categorized consistent symmetry algebras, we look for $P(X,\phi)$ Lagrangians on which these algebras can be realized. In doing this, it is important to keep in mind that we have the freedom of performing field redefinitions $\phi\to \chi(\phi)$. We will use this freedom to bring the putative symmetry transformations in a canonical simple form.

 Let us return to the question of the dependence of $f_{\mu_1..\mu_n}$ on higher derivatives of $\phi$. For concreteness consider a symmetry $\delta \phi$ that depends on the second but not higher derivatives of $\phi$. We demand that the symmetry leaves the action invariant up to a total derivative:
\be
\label{totd}
 \delta \mathcal{L}= P_\phi\,\delta \phi - 2P_X\,\partial^\mu\phi\, \partial_\mu (\delta \phi)=\partial_\mu\,F^\mu\,,
\ee
where $P_{X}\equiv \pdv{P}{X}$ and $P_\phi \equiv \partial_{\phi}P$. Barring the trivial theories for which $  P_{X}=0 $ for all values of $  X $ and $  \phi $, we have to match the dependence on the third derivatives of $\phi$ on both sides of this equation. This implies 
\be
F^\mu=-2P_X\,\partial^\mu\phi\, \delta\phi+G^\mu (\phi,\partial_\nu\phi)\,,
\ee
where $G^\mu$ are arbitrary functions of only $\phi$ and $\del \phi$. Inserting this into \eqref{totd} tells us that the $ \delta \phi$ transformation must be of the form
\be
\label{symmh}
\delta \phi=\dfrac{\del_\mu\,G^\mu}{2\del_\mu(P_X\,\del^\mu\phi)+P_\phi}\,.
\ee
In the denominator of this expression we recognize the equations of motion. So these symmetries are singular on-shell unless $\partial_\mu\,G^\mu$ vanishes on-shell as well. But if that were the case, then we could Taylor expand $  \partial_{\mu}G^{\mu} $ as
\be
\partial_{\mu}G^{\mu}\simeq \left( \text{e.o.m.} \right) g_{1}+\left( \text{e.o.m.} \right)^{2}g_{2}+\dots\,.
\ee 
The only term that could give a symmetry transformation that is non-trivial on shell is the first one. By assumption $  G_{\mu} $ is only a function of $  \phi $ and $  \partial \phi $ but not of $  \partial \partial\phi $. So $  g_{1} $ can also only be a function of $  \phi $ and $  \partial \phi $. Then the dependence of $  \delta \phi $ on $  \partial \partial \phi $ in \eqref{symmh} would cancel out between numerator and denominator and we would find that $  \delta \phi $ does not depend on second derivatives of $  \phi $, contradicting the assumption. We conclude that any symmetry that depends on second derivatives of the field is necessary singular on-shell and, as a result, it does not lead to any new conserved Noether currents. By an analogous argument, symmetries that depend on even higher derivatives of $\phi$ are plagued by the same problem. Thus, in the remainder of this work we consider only symmetries that involve at most one derivative per field. Notice that this conclusion relies on the technical assumption that the symmetry transformation does not contains an arbitrarily large number of derivatives of $  \phi $.

\subsection{Degree zero} 
\label{sub:zeroth_order}
We start by studying generators commuting with spacetime translations
\begin{equation}
	\comm{P_{\mu}}{S_m^0}=0\,.
\end{equation}
The most generic scalar symmetry of that type, namely with $m=0$, can have the following form
\begin{equation}
\delta \phi=g(\phi,X)\,,
\end{equation}
with $g$ an unspecified function. Under this symmetry, the transformation of the Lagrangian should match a total derivative term. Therefore, 
\begin{equation}
-2P_X\partial^\mu\phi \Big(g_{\phi}\partial_\mu\phi+g_X\partial_\mu X\Big)+P_\phi\, g(\phi,X)=\partial_\mu \Big(\partial^\mu \phi\, h_1(\phi,X)+\partial^\mu X\, h_2(\phi,X)\Big)\,,
\end{equation}
where on the RHS we have inserted the most general total derivative term that is consistent with the structure of indices. Notice that the terms $\Box{\phi}\,h_1(\phi,X)$ and $\Box{X}\,h_2(\phi,X)$ on the right-hand side do not have any counterpart on the left-hand side. Hence, the only possible degree zero scalar symmetry has $h_1=h_2=0$ and $g_{\phi}=g_{X}=P_{\phi}=0$, namely a shift symmetry, $\delta\phi=$ const. Implementing a shift-symmetry would mean studying superfluid theories of the form $P(X)$. We leave this case aside since it has been studied extensively in \cite{Pajer:2018egx}. Moving on to vector generators with $m=1$, the only possible symmetry transformation is:
\begin{equation}
	\label{eq:cus}
	\delta_{V^{\mu}(f)}\phi=f(\phi)\partial_{\mu}\phi\,.
\end{equation}
For any differentiable function $f(\phi)$, these generators leave invariant the following action
\begin{equation}
 	\label{eq:general_Cuscuton}
 	\highlight{S=\mathlarger{\int} \d^D x \left(\sqrt{X}-V(\phi)\right)}\,.
 \end{equation} 
for any potential $V(\phi)$. This is the action of the \textit{Cuscuton} field \cite{Afshordi:2006ad}, which has been studied in the context of Dark Energy in \cite{Afshordi:2007yx,Afshordi:2009tt,Gomes:2017tzd,Ito:2019ztb} and of inflation in \cite{Ito:2019fie}. Some more formal properties of the Cuscuton were discussed in \cite{Bhattacharyya:2016mah,Gomes:2017tzd}. Cuscuton perturbations have an infinite speed of sound around homogeneous backgrounds of the form $\phi=\phi(t)$, but the theory is nevertheless causal because the field is not dynamical. Indeed, expanding around such backgrounds one sees that the $\dot\pi^2$ term vanishes: a sign that the Cauchy problem is not well-posed \cite{Bruneton:2006gf,Chagoya:2016inc}. In \cite{Pajer:2018egx} it was pointed out that the shift-symmetric Cuscuton possesses and infinite set of symmetries of the form (\ref{eq:cus}). Here we see that this is still true in the non-shift symmetric case of the action (\ref{eq:general_Cuscuton}) for an arbitrary potential $  V(\phi) $.\\

It remains to show that there are no symmetry generators $S^0_m$ for $m\geq 2$, except for the trivial cases in which these are just proportional to the generators we already found as for example in $  S^{0}_{2m}=S^{0}_{0} \eta_{\mu_{1}\nu_{1}}\dots \eta_{\mu_{m}\nu_{m}} $. One such non-trivial symmetry would take the form
\be
\label{S_m}
\delta_{\mu_1...\mu_m}\phi=g_m(\phi,X)\partial_{\mu_1}\phi\,...\,\partial_{\mu_m}\phi\,.
\ee
Here we have not included terms that contain $\eta_{\mu_i\mu_j}$, e.g. $g_{m-2}(\phi,X)\eta_{\mu_1\mu_2}\partial_{\mu_3}\phi...$, because the structure of indices enforces those terms to cancel separately in the variation of the action\footnote{For example, $  \eta_{\mu\nu} $ vanishes for $  \mu\neq \nu $ but one can certainly find profiles $  \phi(x,t) $ for which $  \partial_{\mu}\phi \partial_{\nu}\phi$ does not.}. Intuitively, those terms would genuinely be $S_{m-2}$ symmetries multiplied by the Minkowski metric. Under the above transformation \eqref{S_m}, the variation of the Lagrangian is
\begin{align}
\label{Lagvar}
&-2P_X\,\partial^\alpha\phi\, \partial_\alpha \left[ g_{m}(\phi,X)\partial_{\mu_1}\phi\,...\,\partial_{\mu_m}\phi \right]+P_\phi\, g_{m}(\phi,X)\partial_{\mu_1}\phi\,...\,\partial_{\mu_m}\phi=\\ \nonumber
& \left( -2\frac{\partial g_m}{\partial\phi}\, X\, P_X-2\,\frac{\partial g_m}{\partial X}\,P_X\,\partial^\alpha\phi\,\partial_\alpha X \right)\,\partial_{\mu_1}\phi\,...\,\partial_{\mu_m}\phi\\ \nonumber 
&-2\,P_X\,g_m\,\partial^\alpha\phi\,\Big(\partial_\alpha\partial_{\mu_1}\phi\,...\,\partial_{\mu_m}\phi+\,...\,+\partial_{\mu_1}\phi\,...\,\partial_\alpha\partial_{\mu_m}\phi\Big)-P_\phi\,g_{m}(\phi,X)\partial_{\mu_1}\phi\,...\,\partial_{\mu_m}\phi \,.
\end{align}
For this transformation to be a symmetry, this variation should be equal to the divergence of some tensor with rank $(1,m)$, i.e.~
\begin{align}
\partial_\alpha\, F^\alpha_{\mu_1...\mu_m}=& \partial_\alpha\, \Big[ h_1 \, \partial^\alpha\phi \partial_{\mu_{1}}\phi\dots\partial_{\mu_{m}}\phi+h_2\,\sum\limits_{\text{perms}}\, \delta^\alpha_{\,\mu_1}\partial_{\mu_2}\phi \dots\partial_{\mu_{\tilde m}}\phi \eta_{\mu_{\tilde m+1}\mu_{\tilde m+2}}\dots+  \\
&\quad  +\sum\limits_{\text{perms}} h_3\delta^\alpha_{\mu_1}\eta_{\mu_2\mu_3}..\eta_{\mu_{m-1}\mu_{m}}\Big]\,,
\end{align}
where $  h_{1,2,3} $ are generic functions of $  \phi $ and $  X $.
Above, we have schematically included representatives of all possible index combinations: for instance, the $  h_{2} $ term stands for terms that involve some number of metric factors $\eta_{\mu_i\mu_j}$, e.g. $h_2\, \delta^\alpha_{\,\mu_1}\partial_{\mu_2}\phi\eta_{\mu_3\mu_4}...$). Notice also that the contribution proportional to $  h_{3} $ is allowed only if $m$ is odd. Just like the $m=0$ case, the term proportional to $h_1$ will generate a $\Box \phi$ operator that does not exist in \eqref{Lagvar}, hence $h_1=0$. The $h_2$ term generates index structures like $\partial_{\mu_i}\partial_{\mu_j}\phi$, while the last term, $h_3$, generates structures like $\partial_{\mu_1}\phi\,\eta_{\mu_2\mu_3}..$. None of these index structures have any counterpart in \eqref{Lagvar}. As a result, we find $h_2=h_3=0$ which then enforces $g_m$ to vanish, i.e.~no symmetris $S_{m}$ exist for $  m \geq 2 $. This concludes the listing of degree zero symmetries. 



\subsection{Degree one} 
\label{sub:first_order_}
The transformations at this order can schematically be written as
\begin{equation}
	\label{eq:first_order_general_transfo}
	\delta_{S^1_{m}}\phi=f_0(\phi,\partial \phi,\dots)+f_1(\phi,\partial \phi,\dots) x\,,
\end{equation}
where we left the $  m $ Lorentz indices implicit. According to (\ref{eq:ladder_equation}), the commutator with translations must give back either $P_{\mu}$ itself or $V_{\mu}(f)$ (recall that we are excluding shift-symmetric theories as they were already studied in \cite{Pajer:2018egx}). We start by looking at scalar generators with $m=0$. In (\ref{eq:first_order_general_transfo}) we must have
\begin{equation}
f_{1}^{\mu}=f(\phi,X)\partial^{\mu}\phi\,,
\end{equation}
with further non-trivial constraints coming from the commutator of this symmetry with $V_{\mu}(f)$. We only found one scalar generator whose commutation relation with $P_{\mu}$ is 
\begin{align}
\comm{P_{\mu}}{\hat D}=aP_{\mu}\,,
\end{align}
for some constant $a\neq 0$. The generator $  \hat D $ can be written as
\begin{equation}
	\delta_{\hat D}\phi= \Delta \phi - a x^{\mu}\partial_{\mu}\phi\,,
\end{equation}
with $f_0=\Delta $ and $f(\phi,X)=-a$ and $ \Delta \neq 0$ . Without loss of generality we can rescale the generator $  \hat D $ to set $a\equiv -1$. We recognize this generator as the dilation operator with commutation relation
\begin{equation}
	\comm{P_{\mu}}{\hat D}=-P_{\mu}\,.
\end{equation}
Up to field redefinitions, the most general dilation invariant Lagrangian takes the form
\begin{equation}
	\label{eq:scaling_lagrang}
	\highlight{
	S=\mathlarger{\int}\d^D x \left[\phi^{\frac{D}{\Delta}}h\left(\phi^{-2 \frac{\Delta+1}{\Delta}}X\right)\right]}\,,
\end{equation}
where $h(y)$ is an arbitrary differentiable function. This is an infinite family of scale-invariant theories, one for each function $h$. One should keep in mind that we are discussing only classical symmetries, which might be anomalous at the quantum level. Therefore there is no obstacle in having scale invariance but not full conformal invariance. We refer the reader to \cite{Pajer:2018egx} for a detailed discussion of this point. We will come back these theories further in Sec.~\ref{sec:scaling_theories} and study the dynamics of their perturbations. For the case of a weight zero scalar field $\varphi$,\, $\Delta_{\varphi}=0$, the symmetry takes the form
\begin{equation}
	\delta_{\hat D} \varphi=x^{\mu}\partial_{\mu}\varphi\, ,
\end{equation}
and the invariant Lagrangian reduces to the shift-symmetric action (see also \cite{Hellerman:2015nra,Monin:2016jmo,Cuomo:2017vzg,Esposito:2016ria})
\begin{equation}
	\label{eq:weight_zero_scaling/conformal}
	\highlight{S=\mathlarger{\int}\d^D x\, X^{D/2}}\,,
\end{equation}
which is furthermore invariant under the full conformal group SO$(D,2)$. Again, we refer to \cite{Pajer:2018egx} for more details on this theory.\\
We found one more scalar generator of degree $N=1$, denoted $W$, whose commutator with $P_{\mu}$ gives the degree zero vector generator $V_{\mu}(f)$. The corresponding transformation takes the form:
\begin{equation}
	\delta_{W(g)} \phi= g(\phi)+\frac{g'(\phi)}{d}x^{\mu}\partial_{\mu}\phi \,.
\end{equation}
This is in fact another infinite dimensional family of scalar symmetries since $g(\phi)$ is an arbitrary differentiable function. In agreement with what was found in \cite{Pajer:2018egx}, these generators restrict the Lagrangian of (\ref{eq:general_Cuscuton}) to the shift-symmetric Cuscuton action
\begin{equation}
	\label{eq:Cuscuton+shift}
	\highlight{
	S=\mathlarger{\int} \d^D x \,\sqrt{X}}\,,
\end{equation}
Before moving on to higher $m$ generators, notice that when restricted to $g(\phi)=d\,\phi$ the generator $W(g)$ becomes the dilation operator $\hat D$ and one finds a scaling Cuscuton action
\begin{equation}
	\label{eq:Cuscuton+scaling}
	\highlight{
	S=\mathlarger{\int} \d^D x \left(\sqrt{X}+\lambda\phi^{\frac{D}{d}}\right) \quad\quad  D\neq1\, ,
	}
\end{equation}
where the scalar field has scaling dimension $\Delta=d$. In $D=1$ spacetime dimension the Cuscuton term $\sqrt{X}$ is a total derivative so we can ignore it. The commutation relation with $V_\mu$ is in this case:
\begin{equation}
	\comm{V_\mu{(f)}}{\hat D}=-V_\mu{(f+\Delta \phi f'(\phi))}\, .
\end{equation}
These are all the scalar symmetries of degree $N=1$. For $m=1$ generators, it is impossible to match the Lorentz indices on both sides of the commutator with $P_\mu$ to yield $V_\mu$ or $P_\mu$. For $m=2$ generators, in principle one can write down transformations of the field whose commutation with translations gives back $P_\mu$. However we have checked that - except for the Lorentz transformation $\delta_{M^{\mu \nu}}\phi$ - none of these transformations are symmetries of the action (\ref{eq:P(X,phi) action}). On the other hand, we have found an $m=2$ symmetry generator $T_{\mu \nu}$ whose commutation relation with $P_\mu$ gives back $V_\mu(f)$: 
\begin{equation}
	\comm{P_\mu}{T_{\rho \sigma}}= 2 c\,  \eta_{\mu[\rho}V_{\sigma]}(f)\, ,
\end{equation}
with $T_{\mu \nu}=T_{[\mu \nu]}$. Without loss of generality we can set $c=1$ and the associated symmetry transformation reads
\begin{equation}
	\label{eq:rank2_Cuscuton_symm}
	\delta_{T^{\mu \nu}(f)}=f(\phi)\left(x_\mu\partial_\nu\phi-x_\nu\partial_\mu\phi\right)\, ,
\end{equation}
where $f(\phi)$ is again an arbitrary differentiable function. The $P(X,\phi)$ action invariant under this symmetry is again the Cuscuton (\ref{eq:general_Cuscuton}). The Cuscuton algebra is then:
\begin{align}
	\comm{P_\mu}{V_\mu}&=0 \,, && \comm{P_\mu}{W(g)}=V_\mu\left(\frac{g'}{d}\right)\,,\\
	\comm{P_\mu}{T_{\rho \sigma}(f)}&=2\eta_{\mu[\rho}V_{\sigma]}(f)\, , && \comm{V_\mu(f)}{T_{\rho \sigma}(g)}=-2\eta_{\mu[\rho}V_{\sigma]}(f\cdot g)\, , \\ 
	\comm{V_{\mu}(f)}{V_{\nu}(\tilde{f})}&=0\,, && \comm{T_{\mu \nu}(f)}{T^{\rho \sigma}(g)}= 4\delta_{[\mu}^{[\rho}\tensor{T}{_{\nu]}^{\sigma]}}(f\cdot g) \,.
\end{align}
Interestingly the symmetry generators $V_\mu$ and $T_{\mu \nu}$ are both of the form $V_\mu(f)\propto f(\phi)P_\mu\,$ and $T_{\mu \nu}(f)\propto f(\phi)M_{\mu \nu}$. That is they are given by an arbitrary function of the field $\phi$ multiplying the Poincar\'e generators. We shall see in Sec.~\ref{sub:Cuscuton_curved_background} that this is in fact true for any background spacetimes. The Cuscuton action (\ref{eq:general_Cuscuton}) is invariant under symmetries of the form "$f(\phi)\times$background isometries".\\ 
For generators with $m\geq 3$ the commutation with $P_\mu$ in \eqref{eq:ladder_equation} returns an $S^0_{m+1}$ generator. But by the argument of Sec. \ref{sub:zeroth_order} there are no $S^0_n$ for $n\geq2$ generators and since the $\delta_{S^1_m}\phi$ transformation can only depend on one $x_\mu$ term we can see that such generator $S^1_m$ must take the form of an $S^1_1$ or $S^1_0$ generator multiplied by the appropriate factor of $\eta$'s.
Since we have already classified such generators we conclude that we have all possible $N=1$ symmetry generators.

\subsection{Degree two} 
\label{sub:second_order_}

By the closure of the algebra, the commutator of any $N=2$ symmetry with translations must give either a dilation $\hat D$ or a Lorentz transformation. The only other possibility for the commutator would be to return $W(g)$ or $T_{\mu \nu}$ but the action (\ref{eq:Cuscuton+shift}) has no more free parameters and so it cannot be constrained further.\footnote{It may be that (\ref{eq:Cuscuton+shift}) has an even larger symmetry algebra, but here our goal is to find all possible symmetric Lagrangians as opposed to all possible symmetry transformations.}  In this case we found two $m=1$ vector symmetry generators. The first one corresponds to the generator of special conformal transformations, for non-zero weight $\Delta\neq 0$ it reads
\begin{equation}
	\delta_{K_{\mu}}\phi=2\Delta \phi x_{\mu} + 2 x_{\mu}x^{\nu}\partial_{\nu}\phi-x^2\partial_{\mu}\phi\,.
\end{equation}
The corresponding invariant action is\footnote{Notice that in the $  D=2 $ case the value of the coupling $  \lambda $ can be set to unity by a linear field redefinition $  \phi \to \phi \lambda^{-\Delta/2}  $.}
\begin{equation}
	\label{eq:5_K_invariant_actions}
	\highlight{
	S=
	\begin{cases}
		\mathlarger{\int} \d^D x \left(X - \lambda\phi^{\frac{2D}{D-2}}\right) & D\neq2 \quad \text{and}\quad \Delta=\frac{D-2}{2}\,,\\
		\mathlarger{\int} \d^D x \left(\frac{X}{\phi^2}- \lambda\phi^{2/\Delta}\right) & D=2\,.
	\end{cases}}
\end{equation}
This situation is different from the shift-symmetric case where there exists no non-trivial conformally invariant action for fields with non-zero weight \cite{Pajer:2018egx}. In $D=4$, we recover the usual $\lambda\phi^4$ conformal theory. In the $D=2$ action, one could make a field redefinition $\phi=e^\chi$ to get a canonical kinetic term in (\ref{eq:5_K_invariant_actions}) but then the scaling symmetry would be less obvious so we preferred to leave the action as is. 
 For fields with weight $\Delta_{\varphi}=0$, the special conformal transformations read
\begin{equation}
	\delta_{K_{\mu}}\varphi= 2x_{\mu}x^{\nu}\partial_{\nu}\varphi- x^2\partial_{\mu}\varphi\,,
\end{equation}
and the invariant action is again (\ref{eq:weight_zero_scaling/conformal}).\\
The second symmetry of degree $N=2$ that we found has also a vector generator:
\begin{equation}
	\label{eq:non_linear_conformal_transfo}
	\delta_{A_{\mu}^{\pm}}\phi= \frac{1}{R_D}(2 R_D\, x_{\mu} - 2 x_{\mu}x^{\nu}\partial_{\nu}\phi+ x^2\partial_{\mu}\phi)\mp R_D (e^{2\phi/R_D}-1)\, \partial_{\mu}\phi\,,
\end{equation}
where $R_D$ is some constant of dimension of length. The corresponding invariant action is 
\begin{equation}
	\label{eq:warped_AdS_action}
	\highlight{
	S^{(\pm)}=\mathlarger{\int}\d^D x \left(e^{-D \phi/R_D}\sqrt{1\pm e^{2\phi/R_D}X}+ \lambda e^{-D \phi/R_D}\right)}\,.
\end{equation}
The minus sign in the square root (and the corresponding plus sign in the transformation) represents the action for a flat D-brane living in an $AdS_{D+1}$ spacetime \cite{Silverstein:2003hf} with radius $R_D$. In this case $\phi$ non-linearly realizes the Conformal algebra $\so(2,D)$ (the isometry group of $AdS_{D+1}$) where the generator $A_{\mu}^{+}$ of (\ref{eq:non_linear_conformal_transfo}) is the generalization of rotation in the extra dimension. The action \eqref{eq:warped_AdS_action} is also invariant under non-linearly realized dilation
\be
\delta_D\phi=-R_D+x^\mu\partial_\mu\phi\,.
\ee
In \eqref{eq:warped_AdS_action}, the plus sign can also be interpreted as a brane embedding but this time in an ambient de Sitter space with two time directions \cite{Goon:2011qf}. Because of this, perturbations around homogeneous backgrounds will generically have super-luminal speed of propagation. More generally this theory will fail to satisfy positivity bounds coming from unitarity of the UV completion \cite{Adams:2006sv}. We will review later in Sec.~\ref{sec:(A)dS_backgrounds} that both these theories have known extensions to (A)dS backgrounds corresponding to different brane geometries and embeddings.\\
When considering the higher-derivatives terms invariant under the symmetry (\ref{eq:non_linear_conformal_transfo}), which were first computed in \cite{deRham:2010eu}, it is interesting to note that the actions (\ref{eq:5_K_invariant_actions}) and (\ref{eq:warped_AdS_action}) are in fact related by a non-linear, field-dependent change of coordinates and field redefinition \cite{Bellucci:2002ji}. They correspond in fact to the non-linear realization of a different basis for the conformal algebra and thus encode the same physics \cite{Creminelli:2013ygt}. However for the truncation to leading-order in derivatives that we consider here, they are two distinct theories\footnote{We thank David Stefanyszyn for a discussion of this point.}.\\
Following arguments similar to those of sections \ref{sub:zeroth_order} and \ref{sub:first_order_} one can show that there are no $S^2_m$ symmetry generator for $m\geq 2$, thus this completes the list of possibilities for $N=2$ generators.


\subsection{Degree three and higher} 
\label{sub:paragraph_name}
We have seen that the generators of order $N=2$ have commutation relations with $P_{\mu}$ of the form
\begin{equation}
	\comm{K_{\mu}}{P_{\nu}}= 2 \left(M_{\mu \nu} - \eta_{\mu \nu} \hat D\right)\,,
\end{equation}
where $\hat D$ is the generator of dilations. Then let us assume an order $N=3$ symmetry generator existed. Its commutator with $P_{\mu}$ would yield an order $N=2$ generator, that is either $K_{\mu}$ or $A_{\mu}$. Since the warped DBI algebra involving $  A_{\mu} $ is the same as the conformal algebra involving $  K_{\mu} $, namely $  \so(2,D) $ without loss of generality we can limit ourselves to just consider $K_{\mu}$. The commutation relation can be written in complete generality as:
\begin{equation}
	\comm{P_{\mu}}{S_m^3}=t_{\alpha \mu \sigma}K^{\sigma}+S^1_{m+1}\,,
\end{equation}
where we followed the same notation as \cite{Pajer:2018egx}: $\alpha=\{\alpha_1,\dots,\alpha_m\}$ and $S_{m+1}^1$ is an order 1 or 0 generator (Dilation, Lorentz transformations or Translations). Then since the structure is that of the conformal algebra as in \cite{Pajer:2018egx}, the theorem proved there applies here too. That is, in $D\geq 3$ the tensors $t_{\alpha\mu\sigma}$ must vanish and thus $  S^{3}_{m} $ is not a degree three symmetry, contradicting out starting assumption. As noted in \cite{Pajer:2018egx} and as well-known from CFT, the case $D=2$ is special: the above theorem does not apply and there exist degree 3 symmetry generators. However the only invariant Lagrangians are those of (\ref{eq:5_K_invariant_actions}) and (\ref{eq:warped_AdS_action}), which have no free parameters and therefore cannot be constrained further. We conclude that there cannot be any new symmetric Lagrangians in any number of spacetime dimensions $  D\geq 2 $.\\
Our classification of symmetric $P(X,\phi)$ theories on flat space is thus complete. The theories listed in Table \ref{tab:all_algebras_mink} exhaust all possibilities for enhanced continuous symmetries of Poincar\'e-invariant scalar fields theories to leading order in derivatives. In the rest of this paper we discuss some aspects of Scaling theories (\ref{eq:scaling_lagrang}) and comment on the extension of our classification to cosmological backgrounds.





\section{Scaling theories on Minkowski spacetime} 
\label{sec:scaling_theories}
In this section, we discuss the scaling-invariant theories in more detail. We first look at the dynamics of perturbations around time-dependent backgrounds. In particular, we derive constraints on the arbitrary function $h(X)$ in (\ref{eq:scaling_lagrang}) that guarantees a healthy dynamics for the fluctuations. Then we look at the implications of scale invariance for perturbations that non-linearly realize broken time-translational invariance. The additional dilation invariance constrains the coefficients of this EFT to satisfy infinitely many recursion relations, which we derive. These results can be interpreted as constraints on the flat-space, decoupling limit of the EFT of Inflation \cite{Cheung:2007st}.

\subsection{Dynamics of perturbations} 
\label{sub:perturbations_around_homogeneous_backgrounds}
Consider a homogeneous background solution $\phi=\bar\phi(t)$. To expand the action (\ref{eq:scaling_lagrang}) around this background, we find it easier to make a field redefinition:
\begin{equation}
	\varphi=\frac{- \Delta}{\phi^{1/\Delta}}\quad\Rightarrow\quad X_{\varphi}= \frac{X}{\phi^{2(\frac{\Delta+1}{\Delta})}}\,.
\end{equation} 
Note that we assume $\Delta\neq0$, otherwise the invariant theory is automatically shift-symmetric. In this case $\varphi$ has scaling dimension $\Delta_{\varphi}=-1$ and the action takes the simpler form
\begin{equation}
	\label{eq:scaling_lagrang_2}
	S=\int\d^D x \frac{h(X)}{\varphi^D}\,.
\end{equation}
We rescale $\varphi$ to make it dimensionless with an arbitrary energy scale $\mu$:
\begin{equation}
	\label{eq:scaling_lagrang_2_dimensionless}
	S=\int\d^Dx \left(\frac{\mu}{\varphi}\right)^D h\left(X/\mu^2\right)\,.
\end{equation}
We now expand around a homogeneous solution $\varphi(x,t)=\bar \varphi(t) +\pi(x,t)$ where $\bar \varphi(t)$ is a solution of the equation of motion:
\begin{equation}
	\label{eq:eom_varphi}
	\frac{\Box \varphi}{\mu^2} h'(X/\mu^2)+\frac{1}{\mu^4}\partial^{\mu}\varphi \partial_{\mu}X h''(X/\mu^2)+\frac{D}{\varphi}\left[\frac{X}{\mu^2}h'(X/\mu^2)-\frac{1}{2}h(X/\mu^2)\right]=0\,.	
\end{equation}
Using this, we can show that the action for $\pi(x,t)$ starts at the quadratic level as it should. The quadratic action is:
\begin{equation}
	S^{(2)}= \int\d^D x \left(\frac{\mu}{\bar\varphi}\right)^D\left[\frac{1}{2}D(D+1)\frac{\bar h}{\bar\varphi^2}\pi^2-2 D \frac{\dbphi}{\bar\varphi}\frac{\bar h'}{\mu^2}\pi\dot\pi+\frac{\dot\pi^2}{\mu^2}\left(\bar h'+2\bar h'' \frac{\dbphi^2}{\mu^2}\right)-\frac{(\nabla\pi^2)}{\mu^2}\bar h'\right]\,,
\end{equation}
where we denoted $\bar h=h(\dot{\bar \varphi}^2/\mu^2),\, \bar h'=h'(X)\Big|_{X=\dot{\bar \varphi}^2/\mu^2}, \, \text{etc}$.
We can integrate by parts the $\pi\dot\pi$ term to get another contribution to the mass $m^2$ of the field. We also note that the kinetic term $\dot\pi^2$ vanishes when
\begin{equation}
	N \equiv \bar h'+2\bar h'' \frac{\dbphi^2}{\mu^2}= 0 \,.
\end{equation}
The only action for which this term always vanishes irrespectively of the background $  \bar \varphi(t) $ is the Cuscuton (\ref{eq:general_Cuscuton}). From now on we assume that the kinetic term is non-zero $N\neq0$. Furthermore we see that to avoid ghost instabilities we need to have
\begin{equation}
	\label{eq:ghost_condition}
	\frac{N}{\bar\varphi^D}\geq 0 \quad \text{(no ghosts)} \,.
\end{equation}
Once this is satisfied we can then rescale the field to make the action canonically normalized
integrating by parts again a $\pi_c\dot\pi_c$ term generated in this transformations the field $\pi_c$ gets another contribution to its time dependent mass. We find the canonically normalized action:
\begin{equation}
	\label{eq:scaling_canonic_pert_action}
	S^{(2)}=\int\d^D x \left[\frac{1}{2}\dot\pi_c^2- \frac{c_s^2}{2}(\nabla\pi_c)^2- \frac{1}{2}m^2\pi_c^2\right]\,,
\end{equation}
with the time-dependent speed of sound given by
\begin{align}
	c_s^2&=\frac{\bar h'}{N}=\frac{\bar h'}{\bar h'+2\bar h'' \frac{\dbphi^2}{\mu^2}}\,,
\end{align}
and the time-dependent mass by
\begin{align}
	\label{eq:time_dependent_mass}
	\frac{1}{2}m^2&=\mu^2 \frac{D}{\bar\varphi^2}\frac{1}{N}\left[-\frac{\bar h}{2}+\frac{\dbphi^2}{\mu^2}\bar h'\right] +\frac{\bar\varphi^D}{N}\left[\frac{\d}{\d t}{\left(\sqrt{\frac{N}{\bar{\varphi}^{D}}}\right)}-\frac{1}{2}\frac{\d^2}{\d t^2}{\left(\frac{N}{\bar{\varphi}^{D}}\right)}\right]\,,
\end{align}
where we used the equation of motion to simplify the first bracket. From the action (\ref{eq:scaling_canonic_pert_action}) we can read off the conditions to avoid gradient instabilities, superluminal speed of propagation and tachyonic masses. These are respectively:
\begin{equation}
		\label{eq:gradient_insta_condition}
		\frac{\bar h'}{\bar h'+2\bar h'' \frac{\dbphi^2}{\mu^2}}\geq0 \quad \text{(no grad. instabilities)}\, ,
\end{equation}
\begin{equation}
		\label{eq:superluminal_condition}
		\bar h''\geq0 \quad \text{(no superluminality)}\, ,
\end{equation}
\begin{equation}
		\label{eq:tachyon_condition}
		\mu^2 \frac{D}{\bar\varphi^2}\frac{1}{N}\left[-\frac{\bar h}{2}+\frac{\dbphi^2}{\mu^2}\bar h'\right] +\frac{\bar\varphi^D}{N}\left[\frac{\d}{\d t}{\left(\sqrt{\frac{N}{\bar{\varphi}^{D}}}\right)}-\frac{1}{2}\frac{\d^2}{\d t^2}{\left(\frac{N}{\bar{\varphi}^{D}}\right)}\right]\geq0 \quad \text{(no tachyons)}\, .
\end{equation}
Notice that in even spacetime dimensions, the condition (\ref{eq:ghost_condition}) implies $\bar h'+2\bar h'' \frac{\dbphi^2}{\mu^2}>0$, together with the gradient instability condition (\ref{eq:gradient_insta_condition}) and the superluminal condition (\ref{eq:superluminal_condition}) this combines  to:
\begin{equation}
	h'(\dbphi^2/\mu^2)>0\quad \quad \text{and} \quad \quad h''(\dbphi^2/\mu^2)\geq0 \,.
\end{equation}
These are in fact necessary conditions for the well-posedness of the Cauchy problem for the scalar waves propagation on this background \cite{Bruneton:2006gf}.
This is all we can say about these theories in complete generality. There is however a specific background profile one could be interested in:
\begin{equation}
	\label{eq:ssp_profile}
 	\bar\varphi(t)=-\mu (t+c)\, ,
 \end{equation} 
where $c$ is some constant. In this case although the background spontaneously breaks time translations, there exist a diagonal combination of scaling and time translations that remains unbroken. The diagonal symmetry evolves the field is in the direction of the background evolution, such that the system is effectively time-independent. Indeed in this case scaling transformations act as time translations,
\begin{equation}
	\delta_{\hat D}\varphi=-\bar\varphi-t\dbphi=c\dbphi=-c \mu\, .
\end{equation}
This is the phenomenon of Spontaneous Symmetry Probing \cite{Nicolis:2011pv}. In  \cite{Nicolis:2011pv}, the authors have also shown that, as long as the symmetry generator commutes with spatial translations, perturbations around the background evolution obey a Goldstone theorem keeping them massless. However this is not the case here, as one can check by plugging (\ref{eq:ssp_profile}) into (\ref{eq:time_dependent_mass}). This is because the dilation operator does not commute with space and time translations so one cannot meaningfully define a spectrum of operators for the would-be Hamiltonian of the effective time-translational invariant system. Thus the background is truly time-dependent and there are no massless Goldstone bosons.


\subsection{Scaling symmetry and the EFT of Inflation} 
\label{sub:effective_theory_of_non_gravitational_scaling_theories}

As an application of our results to Cosmology, we study scaling symmetry in the context of the EFT of Inflation \cite{Cheung:2007st}. Indeed it is well-known that the $P(X,\phi)$ theories (\ref{eq:P(X,phi) action}) we are studying in this paper when coupled to gravity and expanded around a homogeneous background solution can also be described by the decoupling limit of the EFT for Inflation. This motivates us to study the effects of a scaling symmetry on the EFT expansion. Such a study has been done for instance in \cite{Finelli:2018upr} for a shift symmetry. In particular it was shown there that imposing an exact shift symmetry yields an infinite tower of recursion relations among the Wilsonian coefficients and their time-dependence. Furthermore, the same recursion relations continue to apply unchanged even when the theory is expanded around curved FLRW backgrounds. In this section, we derive similar recursions relations among EFT coefficients as implied by the scaling symmetry, while remaining in flat spacetime. We discuss Scaling theories on curved backgrounds in Sec.~\ref{sec:scaling_theories_on_curved_backgrounds}.\\
The flat-space limit of the EFT of Inflation can be simply constructed by considering a non-gravitational Poincar\'e-invariant field theory of a single scalar degree of freedom together with a background configuration that spontaneously breaks time translations and boosts. The form of the EFT is then simply dictated by the non-linear realizations of the broken symmetries. Following the notation of \cite{Finelli:2018upr}, to first order in derivatives of the field, the action for perturbations $\pi(x,t)$ can be organized as:
\begin{equation}
	\label{eq:flat_space_EFT_general_lagrang}
	S=\int\d^Dx \sum_{n=1}^{\infty}\frac{d_n(t+\pi)}{n!}\left(-2\dot\pi+\partial_\mu\pi\partial^\mu\pi\right)^n\, .
\end{equation}
 In terms of the action (\ref{eq:P(X,phi) action}) the above coefficients read
\begin{equation}	
	\label{eq:coef_EFT}
	d_n(t+\pi)=\pdv[n]{P}{X}\Big|_{\phi=\bar\phi(t+\pi)}\dot{\bar\phi}^{2n}(t+\pi) \, .
\end{equation}
We now impose invariance under dilations
\begin{equation}
	\delta_{\hat D}\phi=\Delta\phi+x^{\mu}\partial_\mu\phi \, ,
\end{equation}
with arbitrary weight $\Delta$. In terms of $\pi$ the symmetry acts non-linearly as:
\begin{equation}
	\label{eq:non_linear_pi_scaling}
	\delta_{\hat D}\pi=\Delta \frac{\bar\phi(t+\pi)}{\dot{\bar\phi}(t+\pi)}+t+x^\mu\partial_\mu\pi\, .
\end{equation}
Imposing invariance of the action (\ref{eq:flat_space_EFT_general_lagrang}) up to a boundary term under (\ref{eq:non_linear_pi_scaling}) requires the following recursion relations to be satisfied:
\begin{equation}
	\label{eq:scaling_EFT_recursion_relations}
	\boxed{\left(\Delta+1- \Delta \frac{\bar\phi\ddot{\bar\phi}}{\dot{\bar\phi}^2}\right)(2nd_n+2d_{n+1})=D d_n - \Delta\frac{\bar\phi}{\dot{\bar\phi}}\dot d_n}\, .
\end{equation}
Formally this equation is equivalent to having on-shell current conservation. For fields with zero weight, $\Delta=0$, the recursion relations becomes simply $d_{n+1}=\left(\frac{D}{2}-n\right)d_n$ which agrees with the fact that in this case the action is $P(X,\phi)=X^{D/2}$ (\ref{eq:weight_zero_scaling/conformal}) with background $\bar\phi(t)=\mu t$. The relations are also satisfied for any time-dependent background as one can check using the action (\ref{eq:scaling_lagrang_2}) to evaluate (\ref{eq:coef_EFT}).




\section{Curved backgrounds: dS, AdS and FLRW spacetimes} 
\label{sec:(A)dS_backgrounds}
In this last section, we discuss $P(X,\phi)$ theories\footnote{From now on we will write $X=-g^{\mu \nu}\partial_{\mu}\phi\partial_{\nu}\phi$ with the understanding that the metric is either de Sitter, Anti-de Sitter or FLRW.} on FLRW, dS and AdS backgrounds. We do not provide a complete classification of all symmetric actions, but we discuss what form the symmetries in Table \ref{tab:all_algebras_mink} must take on these spacetimes. In other words, there might be other symmetric theories that are unique to (Anti)-de Sitter or FLRW spacetimes with no analogue on Minkowski that  do not appear here. These new potential symmetries would necessarily have to become trivial in the flat-space limit. We should also mention that most of the theories we discuss in the following have already appeared elsewhere in the literature and we provide references when this is the case. We find it nonetheless useful to include them here to provide a self-contained discussion of $P(X,\phi)$ theories.


\subsection{Cuscuton} 
\label{sub:Cuscuton_curved_background}
Let us begin with a minimally coupled Cuscuton, namely the following action 
\be
S=\int \sqrt{-g}\,d^Dx\, \left[(-\dfrac{1}{2}g^{\mu\nu}\partial_\mu\phi \partial_\nu\phi)^{1/2}-V(\phi)\right]\,,
\ee
and inspect its symmetries on top of FLRW and (A)dS spacetimes. It turns out that the symmetries of the  Cuscuton in flat space generalize to any symmetric space in a simple manner. Consider the metric to have a Killing vector $\xi^\mu$, i.e.~
\be
\nabla_{(\mu}\xi_{\nu)}(x)=0\,.
\ee
Then one can show that the Cuscuton action possesses the following symmetry
\be
\delta_{\xi,f}\phi=f(\phi)\xi^\mu(x)\partial_\mu\phi\,.
\ee
This is an infinite dimensional, field dependent, generalization of the background isometries of \emph{any spacetimes}. The symmetry algebra can be easily derived from the algebra of the background isometries 
\be
[\delta_{\xi_1,f},\delta_{\xi_2,g}]\phi=-f(\phi)g(\phi)[\xi_1,\xi_2]^\mu\partial_\mu\phi\,.
\ee
For instance, on a flat FLRW the new symmetries are given by
\begin{align}
\delta_{V_i(f)}\phi=f(\phi)\partial_i \phi\,, \quad \text{and}\quad \delta_{T_{ij}(f)}\phi=f(\phi)(x^i\partial_j-x^j\partial_i)\phi\,.
\end{align}
For $dS_D$ spacetime it is easiest to express the symmetry group in terms of the coordinates of an ambient flat space ($X^a, a=-1,0,1,..,d$), inside which the dS spacetime is identified with the following hypersurface
\be
\eta_{ab}X^a\,X^b=L_D^2\,.
\ee
Then the $dS_D$ isometries match the Lorentz transformations of the ambient space, namely $M_{ab}$'s, and the Cuscuton action possesses an infinite set of new symmetries given by
\be
\delta_{\tilde{M},f}\phi=f(\phi)(X^a\partial_b-X^b\partial_a)\phi\,.
\ee
As anticipated, if we send the radius $L_D$ of the dS space to infinity, the isometry group SO(1,4) contracts to the Poincare group ISO(1,3) and we recover the Cuscuton theory on flat space with the corresponding symmetries presented in \eqref{eq:cus} and \eqref{eq:rank2_Cuscuton_symm}. 

\subsection{Dirac-Born-Infeld} 
\label{sub:brane_embeddings}
As we noted in our earlier discussion, the DBI and warped DBI Lagrangians correspond to the action of a 4-dimensional Minkowski brane embedded in 5-dimensional Minkowski or AdS spacetime, respectively. The presence of the brane in the higher dimensional spacetime spontaneously breaks the translations of the higher isometry group and the position of the brane acts as the Goldstone mode of the spontaneously broken symmetry. The resulting Lagrangian of the field $\pi$ non-linearly realizes the broken symmetries in the same way as (\ref{eq:warped_AdS_action}). This suggests looking for 4-dimensional (A)dS branes in 5 dimensional Minkowski and AdS/dS spacetimes to find the corresponding actions on (A)dS space. Fortunately, these actions were worked out in a systematic way in \cite{Goon:2011qf}. However, the case of AdS$ _D$ branes is slightly different since the only maximally symmetric spacetime, \textit{with only one time direction}, in which it can be embedded is AdS$ _{D+1}$. To embed it in higher dimensional Minkowski and de Sitter spacetime one needs to perform a Wick rotation on one of the coordinates, as for instance the case of the standard embedding of AdS$_D$ in $\R^{d,2}$. These were not considered in \cite{Goon:2011qf}, we worked them out since they are relevant actions and symmetry breaking patterns for our classification. However we can already anticipate that these action are pathological around non-trivial backgrounds and violate positivity bounds. For clarity, we only give the 4-dimensional actions and the interested reader can find the D-dimensional actions in Appendix \ref{sec:maximally_symmetric_branes_constructions_and_contractions} as well as the transformation laws for the symmetry algebras. Note that all these algebras reduce properly in the limit $R,L\to\infty$ to the corresponding extensions of the Poincar\'e algebra of Sec.~\ref{sec:minkowksi_background} and \cite{Pajer:2018egx}.
\subsubsection{De Sitter} 
\label{sub:de_sitter_background}
The DBI actions, to leading order in derivatives, living on 4-dimensional de Sitter space were constructed in \cite{Goon:2011qf}. There the authors derived three different theories corresponding to a dS$_4$ brane living on a dS$_5$, M$_5$ and AdS$_5$ background, respectively called Type I, Type II and Type III dS DBI.\\
The Type I dS DBI action corresponds to the symmetry breaking patter $\so(1,5)\to\so(1,4)$
\begin{equation}
	\Ll_{dS,I}=\frac{L_{5}^4}{L_4^4}\left(\frac{3}{8}\phi- \frac{1}{4}L_{5}\sin\left(\frac{2\phi}{L_{5}}\right)+\frac{1}{32}L_{5}\sin\left(\frac{4\phi}{L_{5}}\right)\right)-\frac{L_{5}^4}{L_4^4}\sin^4\left(\frac{\phi}{L_{5}}\right)\sqrt{1+\frac{(\partial\phi)^2}{\frac{L_5^2}{L_4^2}\sin^2\left(\frac{\pi}{L_{5}}\right)}}\,.
\end{equation}
The type II dS DBI action corresponds to the symmetry breaking pattern $\iso(1,4)\to\so(1,4)$
\begin{equation}
	\Ll_{dS,II}=-\frac{\phi^4}{L_4^4}\sqrt{1+L_4^2\frac{(\partial\phi)^2}{\phi^2}}+\frac{1}{5}\frac{\phi^5}{L_4^4}\,.
\end{equation}
Finally, the Type III dS DBI action corresponds to the symmetry breaking pattern $\so(2,4)\to\so(1,4)$
\begin{align}
	\Ll_{dS,III}=&\frac{R_{5}^4}{L_4^4}\left(\frac{3}{8}\phi- \frac{1}{4}R_{5}\sinh\left(\frac{2\phi}{R_{5}}\right)+\frac{1}{32}R_{5}\sinh\left(\frac{4\phi}{R_{5}}\right)\right)+ \\
	&\quad -\frac{R_5^4}{L_4^4}\sinh^4\left(\frac{\phi}{R_{5}}\right)\sqrt{1+\frac{(\partial\phi)^2}{\frac{R_5^2}{L_4^2}\sinh^2\left(\frac{\phi}{R_{5}}\right)}}\,.
\end{align}
We should also mention that there exists another way to derive the action of the Goldstone boson for this symmetry breaking pattern through the coset construction, as was done in \cite{Hinterbichler:2012mv}.

\subsubsection{Anti-de Sitter} 
\label{sub:anti_de_sitter_background}
Again there are 3 different DBI theories living on AdS$_4$ corresponding to an AdS$_4$ brane living on a dS$_5^{*}$, $\R_{2,3}$ and AdS$_5$ background, respectively called Type I{*}, Type II{*} and Type III AdS DBI. As noted earlier the first two cases are special and were not included in the original study \cite{Goon:2011qf} since there the embedding space needs two timelike directions - indicated here by $*$. For our purposes however this will only amount to a different sign in front of the kinetic term.\\
The Type I{*} AdS DBI action corresponds to the symmetry breaking patter $\so(2,4)\to\so(2,3)$:
\begin{equation}
	\Ll_{AdS, I^{*}}=\frac{L_{5}^4}{R_4^2}\left(\frac{3}{8}\phi+\frac{1}{4}L_{5}\sin\left(\frac{2\phi}{L_{5}}\right)+\frac{1}{32}L_{5}\sin\left(\frac{4\phi}{L_{5}}\right)\right)-\frac{L_5^4}{R_4^4}\cos^4\left(\frac{\phi}{L_{5}}\right)\sqrt{1-\frac{(\partial\phi)^2}{\frac{L_5^2}{R_4^2}\cos^2\left(\frac{\phi}{L_{5}}\right)}}\,.
\end{equation}
The Type II{*} AdS DBI action corresponds to the symmetry breaking patter $\iso(2,3)\to\so(2,3)$:
\begin{equation}
	\Ll_{AdS, II^{*}}=-\frac{\phi^4}{R_4^4}\sqrt{1-R_4^2\frac{(\partial\phi)^2}{\phi^2}}+\frac{1}{5}\frac{\phi^5}{R_4^4}\,.
\end{equation}
Finally, the Type III AdS DBI action also corresponds to the symmetry breaking patter $\so(2,4)\to\so(2,3)$:
\begin{equation}
	\Ll_{AdS, III}=\frac{R_{5}^4}{R_4^4}\left(\frac{3}{8}\phi+\frac{1}{4}R_{5}\sinh\left(\frac{2\phi}{R_{5}}\right)+\frac{1}{32}R_{5}\sinh\left(\frac{4\phi}{R_{5}}\right)\right)-\frac{R_5^4}{R_4^4}\cosh^4\left(\frac{\phi}{R_{5}}\right)\sqrt{1+\frac{(\partial\phi)^2}{\frac{R_5^2}{R_4^2}\cosh^2\left(\frac{\phi}{R_{5}}\right)}}\,.
\end{equation}


\subsection{Scaling theories} 
\label{sec:scaling_theories_on_curved_backgrounds}
We conclude this section by discussing Scaling theories on flat FLRW and de Sitter spacetimes\footnote{The case of Anti-de Sitter space is very similar.}. The metric of a flat FLRW in conformal coordinates is given by
\be 
ds^2=a^2(\eta)\,\Big(-d\eta^2+d\textbf{x}^2\Big)\,,
\ee
Demanding the theory
\be 
S=\int \sqrt{-g}\, P(X,\phi)\,,
\ee
to be invariant under the scaling
\be \label{vic}
\phi(\eta,\textbf{x}) \to \lambda^\Delta\, \phi(\lambda\eta,\lambda\textbf{x})\,,
\ee
enforces $a(\eta)$ to be a power-law, $a(\eta)=\eta^{1/(\epsilon-1)}$ with $\epsilon\in\R$. In particular, matter and radiation domination are specific cases corresponding to $\epsilon=\frac{3}{2}$ and $\epsilon=2$, respectively. Moreover, $P(X,\phi)$ has to take the following form
\be
\label{eq:scaling_theory_curved_background}
P(X,\phi)=\phi^\alpha\, h(\phi^\beta\, X)\,.
\ee
For $  \Delta \neq 0 $, scaling invariance then fixes the exponents to be
\be
\beta=-\dfrac{2\epsilon}{(\epsilon-1)\Delta}-2\,,\qquad \qquad \alpha=\dfrac{4\epsilon}{(\epsilon-1)\Delta}\,.
\ee
The Lagrangian (\ref{eq:scaling_theory_curved_background}) is then the exact equivalent of the theories we studied in section \ref{sec:scaling_theories} for power-law FLRW spacetimes. Notice that in the exact dS case $\epsilon=0$ - after a field redefinition - the theory is exactly shift symmetric, while for other FLRW solutions this is in general not the case. For $  \Delta =0 $, instead we find $  \epsilon=0 $, namely dS spacetime. In this case $  \alpha $ and $  \beta $ are arbitrary. This is to be expected since any covariant field theory in dS should be invariant under dS isometries, which contain the transformation in \eqref{vic} with $  \Delta=0 $, namely a dS dilation. 
Following the logic of section \ref{sub:effective_theory_of_non_gravitational_scaling_theories}, such a symmetry should lead to constraints similar to (\ref{eq:scaling_EFT_recursion_relations}) on the Wilsonian coefficients of the EFT of Inflation.




\section{Discussion and conclusions} 
\label{sec:discussion}
In this paper, we presented a classification of all possible Poincar\'e-invariant scalar field theories of the form $P(X,\phi)$ that are invariant under additional continuous symmetries. In general, such scalar field theories with non-canonical kinetic terms suffer from radiative corrections that change their functional form. Having an enlarged set of symmetries is a desirable feature that can protect a bare Lagrangian against such radiative instabilities\footnote{Of course one needs to check that these symmetries survive at the quantum level, i.e.~that they are not anomalous.}. The most well-known example of this is the Dirac-Born-Infeld theory (\ref{eq:warped_AdS_action}), which enjoys a higher-dimensional boost and scaling symmetry, non-linearly realizing the conformal algebra. We have found that there exist only two othe classes of symmetric theories: the Cuscuton action (\ref{eq:general_Cuscuton}), which possesses an infinite dimensional symmetry algebra and so-called Scaling theories (\ref{eq:scaling_lagrang}), which are invariant under spacetime dilation (these also include fully conformally invariant theories). All these actions can be found in Table \ref{tab:all_algebras_mink} together with the associated symmetry generators and algebras. For all these theories one can consider backgrounds that spontaneously break some or all the symmetry generators, both Poincar\'e and the additional ones. As usual, Poincar\'e symmmetries are recovered in the UV-limit. Also, it should be noted that all the actions we found in this paper are invariant only up to a total derivative term; in other words they are all so-called Wess-Zumino terms. The Noether currents associated with each symmetry can be found in Appendix \ref{app:noether_currents}.

Going beyond these formal considerations and with cosmological applications in mind, we were also interested in studying the resulting dynamical properties of these theories when expanded around homogeneous background field configurations. Again in the context of the shift-symmetric DBI action, the non-linearly realized ISO$(D,1)$ symmetry fully fixes the form of the action for perturbations. In particular all coupling constants to leading order in derivatives are fixed in terms of the speed of sound $c_s$. On the contrary, for the warped DBI theory the dynamics of perturbations is not fully fixed by the symmetry and Wilsonian coefficients are in general time-dependent \cite{Creminelli:2013xfa}. The Cuscuton theory is also special in this regard. While formally the speed of sound of perturbations is infinite, the theory is non-dynamical around cosmological backgrounds: the equation of motion are only first order in time derivatives  (see \cite{Afshordi:2006ad} for further discussion). For this reason we focused our attention in this paper on the dynamics of Scaling theories. We derived in Sec.~\ref{sub:perturbations_around_homogeneous_backgrounds} the dynamics of perturbations together with the conditions needed to avoid gradient instabilities, superluminal propagation and tachyonic masses. However, in this case too the Lagrangian is time-dependent which makes it hard to study the phenomenology in greater depth (see e.g.~\cite{Burgess:2014lwa} for a discussion of this point). As a simple exercise, we also derived the implications of a scaling symmetry in the framework of the EFT of Inflation, obtaining an infinite set of recursion relations amongst the Wilsonian coefficients.\\
Finally we looked at the behaviour of these three classes of theories around Anti-de Sitter, de Sitter or FLRW geometries. It turns out that each class of theory exists on either of these spacetimes. Again the Cuscuton is special: coupled to any background geometry, the action is always invariant under symmetries of the form "free function of $\phi$ $\times$ background isometries". The DBI brane construction also allows for both dS and AdS brane geometries embedded in either dS, AdS or Minkowski space. Finally dilation invariant theories can be easily constructed on power-law FLRW spacetimes, while every covariant theory is scaling invariant on (A)dS spacetimes.\\
There are many of avenues for future work:
\begin{itemize}
\item It would be interesting to extend our classification to the curved backgrounds we studied in Sec.~\ref{sec:(A)dS_backgrounds} and to show whether there exist additional symmetric theories on these backgrounds or not. Since it seems unlikely that a systematic analysis would work for all three spacetimes one might need a more sophisticated approach than the one used in this paper. A recent example of this idea can be found in \cite{Bonifacio:2018zex} where the authors classified (extended) shift-symmetric theories on (A)dS.
\item A second extension of our results would be to higher derivative scalar theories. This would capture for instance Galilean symmetry \cite{Nicolis:2008in}, which only shows up to seconder order in derivatives for interacting theories. This would be particularly relevant in the context of (beyond) Horndeski theories of Dark Energy and Modified Gravity \cite{Horndeski:1974wa,Zumalacarregui:2013pma,Gleyzes:2014dya,Gleyzes:2014qga} in which the $P(X,\phi)$ Lagrangian studied here is only the first term\footnote{For instance in \cite{Padilla:2013jza} the authors derived the most general (globally and locally) scale invariant Horndeski theory.}.
\item The results in this work can presumably be re-phrased in the language of the coset construction and inverse-Higgs constraints. It would be interesting to see if this different but presumably equivalent approach leads to more manageable algebraic manipulations or some additional insight.
\end{itemize}   
Finally, as mentioned in the introduction, a hallmark of flat space scattering amplitudes is the Coleman-Mandula theorem, whereas in cosmology little is known about the wider landscape of possible symmetries that can be obeyed by cosmological correlators. It would be interesting to investigate what restrictions exists for inflationary correlators. The results of this paper are a small step towards this goal and a better understanding of the space of all possible non-linearly realized symmetries.


%
%

\section*{Acknowledgments} 
We thank Garrett Goon, Kurt Hinterbichler, Mehrdad Mirbabayi, Rachel Rosen, David Stefanyszyn and Jakub Supel for stimulating discussions. We are also grateful to David Stefanyszyn for comments on a draft. S.J. and E. P. have been supported in part by the research program VIDI with Project No. 680-47-535, which is (partly) financed by the Netherlands Organisation for Scientific Research (NWO). T.G. is supported by the Cambridge Trust. 
\label{sec:acknowledgments}


\appendix

\section{Symmetry transformations}
\label{app:definition_comm}
Here we review very explicitly how symmetry generators act on the scalar field $\phi(x)$. For a generic symmetry generator $Q$ and a generic operator $\cal O$ we define
\be 
\delta_Q {\cal O}\equiv [Q,{\cal O}]\,,
\ee
and in particular
\be
[Q,\phi(x)]=\delta_Q\,\phi(\phi(x),\partial_\mu\phi(x);x)\,.
\ee
Using the Jacobi identities, it is easy to observe that the algebra of $\delta_{Q}$'s is the same as that of the $Q$'s, 
\be
[[Q_1,Q_2],{\cal O}]=[Q_1,[Q_2,{\cal O}]]-[Q_2,[Q_1,{\cal O}]]=\delta_{Q_1} \delta_{Q_2}{\cal O}-\delta_{Q_2} \delta_{Q_1}{\cal O}\,.
\ee
Therefore if a set of charges $Q_i$ form a Lie algebra, i.e.
\be
[Q_i,Q_j]=c_{ijk}Q_k\,,
\ee
then 
\be
\delta_{Q_i}\delta_{Q_j}{\cal O}-\delta_{Q_j}\delta_{Q_i}{\cal O}=c_{ijk}\delta_{Q_k}{\cal O}\,.
\ee
For an arbitrary operator that depends on $\phi$, its derivatives and $x$, one can show that in the active view 
\be
\delta_Q C=\dfrac{\partial C}{\partial \phi} \delta_Q \phi+\dfrac{\partial C}{\partial (\partial_\mu \phi)}\partial_\mu \delta_Q\phi\,,
\ee
from which it follows that 
\begin{align}
[\delta_{Q_1},\delta_{Q_2}]\phi &=\dfrac{\partial\delta_{Q_2}\phi}{\partial\phi}\delta_{Q_1}\phi-
\dfrac{\partial\delta_{Q_1}\phi}{\partial\phi}\delta_{Q_2}\phi\\ \nonumber
&+\dfrac{\partial \delta_{Q_2}\phi}{\partial\,\partial_\mu\phi}\partial_\mu\delta_{Q_1}\phi
-\dfrac{\partial \delta_{Q_1}\phi}{\partial\,\partial_\mu\phi}\partial_\mu\delta_{Q_2}\phi\,.
\end{align}

\section{Maximally symmetric DBI actions in $D $-dimensions} 
\label{sec:maximally_symmetric_branes_constructions_and_contractions}

In this appendix we list the actions of D-dimensional maximally symmetric branes embedded in (D+1)-dimensional maximally symmetric spaces from \cite{Goon:2011qf} together with three more brane embeddings corresponding to the cases where the embedding space has two time dimensions. We also give the explicit form of the non-linear transformations acting on the scalar field - that we denote $\pi$ here. These provide the extensions of the usual DBI and Warped DBI actions to (Anti-)de Sitter geometries. One can explicitly verify that in the limit where the embedded brane radius, $R_D$ or $L_D$, goes to infinity the action reduces to the action for a Minkowski brane embedded in the corresponding space. At the level of the isometry algebra this is equivalent to a Inon\"u-Wigner contraction.

\subsection{de Sitter Branes} 
\label{sub:de_sitter_branes}

\paragraph{Type I dS DBI} 
\label{subsub:type_i_ds_dbi}
The action for the modulus of a unit radius $dS_D$ brane in a $dS_{D+1}$ spacetime is:
\begin{align}\nonumber
	S=\int\sqrt{-g}\Bigg[&-L_{D+1}^{1+D}\cos\left(\frac{\pi}{L_{D+1}}\right)\pFq{2}{1}\left(\frac{1}{2},\frac{1-D}{2},\frac{3}{2},\cos^2\left(\frac{\pi}{L_{D+1}}\right)\right)\\ 
	&-L_{D+1}^D\sin^D\left(\frac{\pi}{L_{D+1}}\right)\sqrt{1+\frac{(\partial\pi)^2}{L_{D+1}^2\sin^2\left(\frac{\pi}{L_{D+1}}\right)}}\Bigg]
\end{align}
for instance in $D=4$ this gives
\begin{equation}
	S=\int\sqrt{-g}\left[\frac{L_{5}^4}{32}\left(12\pi-8L_{5}\sin\left(\frac{2\pi}{L_{5}}\right)+L_{5}\sin\left(\frac{4\pi}{L_{5}}\right)\right)-L_{5}^4\sin^4\left(\frac{\pi}{L_{5}}\right)\sqrt{1+\frac{(\partial\pi)^2}{L_{5}^2\sin^2\left(\frac{\pi}{L_{5}}\right)}}\right]
\end{equation}
where $L_{D+1}$ is the radius of the (D+1)-dimensional de Sitter space. In Poincar\'e coordinates
\begin{equation}
	\d s^2=\frac{1}{\eta^2}(-\d\eta^2+\d\vec{x}^2)
\end{equation}
the additional generators act non-linearly on the field as follows:
\begin{align}
	\delta_{+}\pi&=\frac{L_{D+1}}{\eta}-\cot\left(\frac{\pi}{L_{D+1}}\right)\pi'\\
	\delta_{-}\pi&=\frac{L_{D+1}}{\eta}(-\eta^2+x^2)-(\eta^2+x^2)\cot\left(\frac{\pi}{L_{D+1}}\right)\pi'-2\eta\cot\left(\frac{\pi}{L_{D+1}}\right)x^i\partial_i\pi\\
	\delta_{i}\pi&= \frac{L_{D+1}}{\eta}x_i-x_i\cot(\frac{\pi}{L_{D+1}})\pi'-\eta\cot\left(\frac{\pi}{L_{D+1}}\right)\partial_i\pi
\end{align}
which realizes the symmetry breaking pattern
\begin{equation}
	\so(D+1,1)\to\so(D,1)
\end{equation}

\paragraph{Type II dS DBI} 
\label{subsub:type_ii_ds_dbi}
The action for the modulus of a unit radius $dS_D$ brane in a $M_{D+1}$ spacetime is:
\begin{equation}
	\label{eq:type_ii_ds_dbi}
	S=\int\sqrt{-g}\left[-\pi^D\sqrt{1+\frac{(\partial\pi)^2}{\pi^2}}+\frac{1}{D+1}\pi^{D+1}\right]
\end{equation}
In Poincar\'e coordinates again, the additional generators act non-linearly on the field as follows:
\begin{align}
	\delta_{+}\pi&=\frac{1}{\eta}(\eta^2-x^2)+\frac{1}{\pi}(\eta^2+x^2)\pi'+2 \frac{\eta}{\pi}x^i\partial_i\pi\\
	\delta_{-}\pi&=-\frac{1}{\eta}+\frac{1}{\pi}\pi'\\
	\delta_{i}\pi&= \frac{1}{\eta}x_i- \frac{x_i}{\pi}\pi'-\frac{\eta}{\pi}\partial_i\pi
\end{align}
which realizes the symmetry breaking pattern
\begin{equation}
	\iso(D,1)\to\so(D,1)
\end{equation}


\paragraph{Type III dS DBI} 
\label{subsub:type_iii_ds_dbi}
The action for the modulus of a unit radius $dS_D$ brane in a $AdS_{D+1}$ spacetime is:
\begin{align}\nonumber
	S=\int\sqrt{-g}\Bigg[&-R_{D+1}^{1+D}\cosh\left(\frac{\pi}{R_{D+1}}\right)\pFq{2}{1}\left(\frac{1}{2},\frac{1-D}{2},\frac{3}{2},\cosh^2\left(\frac{\pi}{R_{D+1}}\right)\right)\\ 
	&-R_{D+1}^D\sinh^D\left(\frac{\pi}{R_{D+1}}\right)\sqrt{1+\frac{(\partial\pi)^2}{R_{D+1}^2\sinh^2\left(\frac{\pi}{R_{D+1}}\right)}}\Bigg]
\end{align}
for instance in $D=4$ this gives
\begin{equation}
	S=\int\sqrt{-g}\left[\frac{R_{5}^4}{32}\left(12\pi-8R_{5}\sinh\left(\frac{2\pi}{R_{5}}\right)+R_{5}\sinh\left(\frac{4\pi}{R_{5}}\right)\right)-R_{5}^4\sinh^4\left(\frac{\pi}{R_{5}}\right)\sqrt{1+\frac{(\partial\pi)^2}{R_{5}^2\sinh^2\left(\frac{\pi}{R_{5}}\right)}}\right]
\end{equation}
where $R_{D+1}$ is the radius of the (D+1)-dimensional de Sitter space. In Poincar\'e coordinates the additional generators act non-linearly on the field as follows:
\begin{align}
	\delta_{+}\pi&=\frac{R_{D+1}}{\eta}-\coth\left(\frac{\pi}{R_{D+1}}\right)\pi'\\
	\delta_{-}\pi&=\frac{R_{D+1}}{\eta}(-\eta^2+x^2)-(\eta^2+x^2)\coth\left(\frac{\pi}{R_{D+1}}\right)\pi'-2\eta\coth\left(\frac{\pi}{R_{D+1}}\right)x^i\partial_i\pi\\
	\delta_{i}\pi&= \frac{R_{D+1}}{\eta}x_i-x_i\coth(\frac{\pi}{R_{D+1}})\pi'-\eta\coth\left(\frac{\pi}{R_{D+1}}\right)\partial_i\pi
\end{align}
which realizes the symmetry breaking pattern
\begin{equation}
	\so(D,2)\to\so(D,1)
\end{equation}
\subsection{Anti-de Sitter Branes} 
\label{sub:anti_de_sitter_branes}
\paragraph{Type I{*} AdS DBI} 
\label{subsub:type_iii_ads_dbi}
This case we want to embed $AdS_D$ into "$dS_{D+1}$" with two time dimensions. To be more precise about what this mean, we can take the usual embedding of de Sitter space in Minkowski space, except that we now Wick rotate one of the spatial coordinates:
\begin{equation}
	-(X^0)^2+\sum_{i=1}^{D}(X^i)^2-(X^{D+1})^2=L_{D+1}^2
\end{equation}
effectively this could also be seen as a Anti-de Sitter space with imaginary radius: $L_{D+1}=i R_{D+1}$. From this observation we can use the embedding of $AdS_D$ into $AdS_{D+1}$ and Wick rotate the (D+1)-dimensional radius. The embedding space metric is then:
\begin{equation}
	\d s^2=\d\rho^2-L_{D+1}^2\cos^2\left(\frac{\rho}{L_{D+1}}\right)\, \d s_{AdS_D}^2
\end{equation}
where $\d s_{AdS_D}^2$ is the $AdS_D$ metric in global coordinates. This gives
\begin{equation}
	f(\pi)=iL_{D+1}\cos\left(\frac{\pi}{L_{D+1}}\right),  \quad \quad g_{\mu \nu}=g_{\mu \nu}^{AdS_D}
\end{equation}
and thus the invariant action is:
\begin{align}\nonumber
	S=\int\sqrt{-g}\Bigg[&-R_{D+1}^{1+D}(1+D)^{-1}\cos^{1+D}\left(\frac{\pi}{R_{D+1}}\right)\pFq{2}{1}\left(\frac{1}{2},\frac{1+D}{2},\frac{3+D}{2},\cos^2\left(\frac{\pi}{R_{D+1}}\right)\right)\\ 
	&-R_{D+1}^D\cos^D\left(\frac{\pi}{R_{D+1}}\right)\sqrt{1-\frac{(\partial\pi)^2}{R_{D+1}^2\cos^2\left(\frac{\pi}{R_{D+1}}\right)}}\Bigg]
\end{align}
for instance in $D=4$ this is:
\begin{align}\nonumber
	S=\int\sqrt{-g}\Bigg[&\frac{L_{D+1}^4}{32}\left(12\pi+8L_{D+1}\sin\left(\frac{2\pi}{L_{D+1}}\right)+L_{D+1}\sin\left(\frac{4\pi}{L_{D+1}}\right)\right)\\ 
	&-L_{D+1}^4\cos^4\left(\frac{\pi}{L_{D+1}}\right)\sqrt{1-\frac{(\partial\pi)^2}{L_{D+1}^2\cos^2\left(\frac{\pi}{L_{D+1}}\right)}}\Bigg]
\end{align}
In D-dimensional Poincar\'e coordinates for Anti-de Sitter space, the non-linearly realized symmetries are:
\begin{align}
	\delta_{+}\pi&=-\frac{L_{D+1}}{z}-\tan\left(\frac{\pi}{L_{D+1}}\right)\pi'\\
	\delta_{-}\pi&=-\frac{L_{D+1}}{z}(z^2+x^2)-(-z^2+x^2)\tan\left(\frac{\pi}{L_{D+1}}\right)\pi'+2z\tan\left(\frac{\pi}{L_{D+1}}\right)x^i\partial_i\pi\\
	\delta_{i}\pi&= -\frac{L_{D+1}}{z}x_i-x_i\tan(\frac{\pi}{L_{D+1}})\pi'+z\tan\left(\frac{\pi}{L_{D+1}}\right)\partial_i\pi
\end{align}

\paragraph{Type II{*} AdS DBI} 
\label{subsub:type_ii_}
This case we embed a D-dimensional Anti-de Sitter spacetime in Lorentzian spacetime $\R^{D-1,2}$:
\begin{equation}
	\d s^2=-(\d X^0)^2+\sum_{i=1}^{D-1}(\d X^i)^2-(\d X^D)^2
\end{equation}
by the codimension one hypersurface:
\begin{equation}
	-(X^0)^2+\sum_{i=1}^{D-1}(X^i)^2-(X^D)^2=-R^2
\end{equation}
We use global coordinates to parametrize the $AdS_d$ metric:
\begin{align}
	X^0&=R\cosh\rho\cos\tau\\
	X^d&=R\cosh\rho\sin\tau\\
	X^i&=R\sinh\rho\, \hat{x}^i \quad \quad \quad i=1,\dots,D-1
\end{align}
where $\sum_{i=1}^{D-1}(\hat{x}^i)^2=1$ such that the D-dimensional metric is
\begin{equation}
	\d s^2=-\d R^2+R^2(-\cosh^2\rho\d\tau^2+\d\rho^2+\sinh^2\rho\d\Omega_{D-2}^2)
\end{equation}
where $\d\Omega_{D-2}^2$ is the round metric on $\mathbb{S}^{D-2}$. In the notation of \cite{Goon:2011qf} this translates to:
\begin{equation}
	f(\pi)=\pi, \quad \quad g_{\mu \nu}= g_{\mu \nu}^{AdS_D}
\end{equation}
However the Gaussian transverse coordinate $R$ is here Wick rotated so we will get a sign difference in the $\gamma$ factor of \cite{Goon:2011qf} compared to the de Sitter brane action. Thus the invariant action is:
\begin{equation}
	S=\int\sqrt{-g}\left[-\pi^D\sqrt{1-\frac{(\partial\pi)^2}{\pi^2}}+\frac{1}{D+1}\pi^{1+D}\right]\,.
\end{equation}
In global coordinates, the additional generators act non-linearly on the field as follows:
\begin{align}
	\delta_{+}\pi&=\cosh\rho (\cos\tau+\sin\tau)+\pi\sinh\rho (\cos\tau+\sin\tau)\partial_{\rho}\pi-\pi\cosh\rho(\sin\tau-\cos\tau)\partial_{\tau}\pi\,,\\
	\delta_{-}\pi&=\cosh\rho (\cos\tau-\sin\tau)+\pi\sinh\rho (\cos\tau-\sin\tau)\partial_{\rho}\pi-\pi\cosh\rho(\sin\tau+\cos\tau)\partial_{\tau}\pi\,,\\
	\delta_{i}\pi&= \hat{x}_i\sinh\rho\, +\hat{x}_i \pi \cosh\rho   \, \partial_{\rho}\pi+\pi\sinh\rho\, \partial_i\pi\,,
\end{align}
which realizes the symmetry breaking pattern
\begin{equation}
	\iso(D-1,2)\to\so(D,2)
\end{equation}

\paragraph{Type III AdS DBI} 
\label{subsub:type_i_ads_dbi}
The action for a unit radius D-dimensional Anti-de Sitter brane embedding in a (D+1)-dimensional Anti-de Sitter spacetime was worked out in \cite{Goon:2011qf}. The result in D-dimensions is
\begin{align}\nonumber
	S=\int\sqrt{-g}\Bigg[&-R_{D+1}^{1+D}(1+D)^{-1}\cosh^{1+D}\left(\frac{\pi}{R_{D+1}}\right)\pFq{2}{1}\left(\frac{1}{2},\frac{1+D}{2},\frac{3+D}{2},\cosh^2\left(\frac{\pi}{R_{D+1}}\right)\right)\\ 
	&-R_{D+1}^D\cosh^D\left(\frac{\pi}{R_{D+1}}\right)\sqrt{1+\frac{(\partial\pi)^2}{R_{D+1}^2\cosh^2\left(\frac{\pi}{R_{D+1}}\right)}}\Bigg]\,.
\end{align}
For instance in $D=4$ this is:
\begin{equation}
	S=\int\sqrt{-g}\left[\frac{R_{5}^4}{32}\left(12\pi+8R_{5}\sinh\left(\frac{2\pi}{R_{5}}\right)+R_{5}\sinh\left(\frac{4\pi}{R_{5}}\right)\right)-R_{5}^4\cosh^4\left(\frac{\pi}{R_{5}}\right)\sqrt{1+\frac{(\partial\pi)^2}{R_{5}^2\cosh^2\left(\frac{\pi}{R_{5}}\right)}}\right]\,,
\end{equation}
where $R_{D+1}$ is the radius of the (D+1)-dimensional Anti-de Sitter space. In $D $-dimensional Poincar\'e coordinates
\begin{equation}
	\d s^2=\frac{1}{z^2}(\d z^2+\eta_{ij}\d x^i\d x^j)\,.
\end{equation}
The additional generators act non-linearly on the field as follows:
\begin{align}
	\delta_{+}\pi&=\frac{R_{D+1}}{z}-\tanh\left(\frac{\pi}{R_{D+1}}\right)\pi'\,,\\
	\delta_{-}\pi&=\frac{R_{D+1}}{z}(z^2+x^2)-(-z^2+x^2)\tanh\left(\frac{\pi}{R_{D+1}}\right)\pi'+2z\tanh\left(\frac{\pi}{R_{D+1}}\right)x^i\partial_i\pi\\
	\delta_{i}\pi&= \frac{R_{D+1}}{z}x_i-x_i\tanh(\frac{\pi}{R_{D+1}})\pi'+z\tanh\left(\frac{\pi}{R_{D+1}}\right)\partial_i\pi\,,
\end{align}
which realizes the symmetry breaking pattern
\begin{equation}
	\so(D,2)\to\so(D-1,2)\,.
\end{equation}

\newpage
\section{Noether currents}
\label{app:noether_currents}
In this appendix we list the Noether currents associated with each new symmetry found in this paper. 
	\subsection*{\underline{$\sqrt{X}+V(\phi)$}} 
	\begin{align}
		J^\mu_{V_\nu}&=-\dfrac{\partial^\mu\phi}{\sqrt{X}}f(\phi)\partial_\nu\phi-\delta^\mu_\nu\,\Big[\sqrt{X}f(\phi)-\dfrac{1}{2}\int d\phi\,V_\phi\,f(\phi)\Big]\, ,\\
		J^\mu_{T_{\alpha\nu}}&=-\dfrac{\partial^\mu\phi}{\sqrt{X}}(x_\alpha\partial_\nu\phi-x_\nu\partial_\alpha\phi)-(f\,\sqrt{X}+\int d\phi\,V_\phi\,f(\phi)\,)\,(x_\alpha\delta^\mu_\nu-x_\nu\delta^\mu_\alpha)\, .
	\end{align}
	\subsection*{\underline{$\sqrt{X}+\lambda\,\phi^{\frac{D}{D-1}}$}}
	\begin{align}
		J_{\hat D}^\mu&= -\dfrac{\partial^\mu\phi}{\sqrt{X}}\Big((D-1)\phi+x^\nu\partial_\nu\phi\Big)-x^\mu\,(\sqrt{X}+\lambda\,\phi^{\frac{D}{D-1}})\, .
	\end{align}
	\subsection*{\underline{$\phi^{D/\Delta}\,h(\phi^{-\frac{2(\Delta+1)}{\Delta}}\,X)$}}
	\begin{align}
		J^\mu_{\hat D}&=-2\partial^\mu\phi\,h'(\phi^{-\frac{2(\Delta+1)}{\Delta}}\,X)\phi^{D/\Delta}\Big(\Delta \phi+x^\nu\partial_\nu\phi\Big) -x^\mu\,\phi^{D/\Delta}\,h(\phi^{-\frac{2(\Delta+1)}{\Delta}}\,X)\, .
	\end{align}
	\subsection*{\underline{$ X +\lambda\, \phi^{\frac{2D}{D-2}}$} ($D\neq 2$)}
	\begin{align}
		J^\mu_{\hat D}=&-2\mu^{D-2}\,\partial^\mu\phi \Big(\frac{D-2}{2}\,\phi+x^\nu\partial_\nu\phi\Big)-\mu^D \,x^\mu\,\Big(-\mu^{-2}\, X+\lambda\,\phi^{\frac{2D}{D-2}}\Big)\, ,\\
J_{K_\nu}^\mu =& -2\mu^{D-2}\partial^\mu\phi\,\Big[(D-2)x_\nu\,\phi+2x_\nu x^\alpha\partial_\alpha\phi-x^2\partial_\nu\phi\Big]\\ \nonumber
&+2\mu^{D-1}\, x_\nu\,x^\mu\,(\partial\phi)^2-\delta^\mu_\nu\, \mu^{D-1}\Big[\,x^2(\partial\phi)^2-\phi^2\Big]-\mu^{D+1}\, \lambda(2x_\nu x^\mu-\delta^\mu_\nu\,x^2)\phi^{\frac{2D}{D-2}}\, .
	\end{align}
	\subsection*{\underline{$\dfrac{X}{\phi^2}+\lambda\, \phi^{2/\Delta}$}($D=2$)}
	\begin{align}
		J^\mu_{\hat D}&=-\dfrac{2\partial^\mu\phi}{\phi^2}(\Delta\,\phi+x^\nu\partial_\nu\phi)-x^\mu\,(\dfrac{X}{\phi^2}+\lambda\,\phi^{2/\Delta})\, ,\\
		J^\mu_{K_\nu}&=-\dfrac{2\partial^\mu\phi}{\phi^2}(2\Delta\,\phi\,x_\nu+2x_\nu\,x^\alpha \partial_\alpha\phi-x^2\partial_\nu\phi)
\\ \nonumber
&+4\Delta\,\ln (\phi)\delta^\mu_\nu-\dfrac{2X}{\phi^2}x^\mu\,x_\nu+\dfrac{X}{\phi^2}x^2\,\delta^\mu_\nu -\lambda\,(2x^\mu\,x_\nu\phi^{2/\Delta}-x^2\,\phi^{2/\Delta}\delta^\mu_\nu)\, .
	\end{align}
	\subsection*{\underline{$e^{-D\phi/R_D} \left( \sqrt{1\pm e^{2\phi/R_D}X}+ \lambda  \right)$}}
	\begin{align}
		J^\mu_{\hat D}&=\mp e^{-(2-D)\phi/R_D}\, (1\pm e^{2\phi/R_D}X)^{-1/2}\,\partial^\mu\phi\, (-R_D+x^\nu\partial_\nu\phi)\\ \nonumber
		 &-x^\mu\, e^{-D\phi/R_D}\Big(\sqrt{1\pm e^{2\phi/R_D}X}+ \lambda \Big)\, ,\\\nonumber
		J^\mu_{A_\nu} &=\mp \dfrac{e^{(2-D)\phi/R_D}}{\sqrt{1\pm e^{2\phi/R_D}}}\Big[2\partial^\mu\phi x_\nu-\frac{2}{R_D}x^\alpha\partial_\alpha\phi \partial^\mu\phi\,x_\nu+\frac{x^2}{R_D}\partial^\mu\phi\,\partial_\nu\phi\mp R_D(e^{2\phi/R_D}-1)\partial^\mu\phi\partial_\nu\phi\Big]\\ \nonumber
&-e^{-D\phi/R_D}\sqrt{1\pm e^{2\phi/R_D}X}\,\Big[-\frac{2}{R_D}x^\mu x_\nu+\delta^\mu_\nu\, \frac{x^2}{R_D}\mp \delta^\mu_\nu\,R_D(e^{2\phi/R_D}-1)\Big]\\
&+\lambda\left\lbrace
\begin{array}{cc}
\delta^\mu_\nu\, \frac{2R_D}{2-D}e^{(2-D)\phi/R_D}& D\neq 2\\ 
&\\
2\phi\, \delta^\mu_\nu & D=2
\end{array}\right . \, .
	\end{align}

\bibliographystyle{JHEP}
\bibliography{symmetric_scalars}

\end{document}